\documentclass{PoS}

\usepackage{graphics}

\usepackage{epsfig}

\def\text{{}}   
\def\lsim{\raise0.3ex\hbox{$<$\kern-0.75em\raise-1.1ex\hbox{$\sim$}}}
\def\gsim{\raise0.3ex\hbox{$>$\kern-0.75em\raise-1.1ex\hbox{$\sim$}}}

\PoS{PoS(LAT2005)192}

\title{Static quark anti-quark free and internal energy in 2-flavor QCD and
  bound states in the QGP}

\ShortTitle{Static quark anti-quark free and internal energy in 2-flavor QCD}

\author{\speaker{Olaf Kaczmarek}\\
   Fakult\"{a}t f\"{u}r Physik,
 Universit\"{a}t Bielefeld, D-33615 Bielefeld, Germany\\
        E-mail: \email{okacz@physik.uni-bielefeld.de}
}

\author{Felix Zantow\\
  Brookhaven National Laboratory, Upton, NY
  11973, USA\\
  E-mail: \email{zantow@quark.phy.bnl.gov}}

\date{\today}

\abstract{
We present results on heavy quark free energies in 2-flavour QCD.
The temperature dependence of the interaction between static quark
anti-quark pairs will be analyzed in terms of temperature dependent screening
radii, which give a first estimate on the
medium modification of (heavy quark) bound states in the quark gluon plasma.
Comparing those radii to the (zero temperature) mean squared charge radii of charmonium
states indicates that the $J/\psi$ may survive the phase transition as a bound
state, while $\chi_c$ and $\psi'$ are expected to show significant thermal
modifications at temperatures close to the transition.
Furthermore we will analyze the relation between heavy quark free energies,
entropy contributions and internal energy and discuss their relation to
potential models used to analyze the melting of heavy quark bound states above
the deconfinement temperature. Results of different groups and various
potential models for bound states in the deconfined phase of QCD are compared. 
}

\FullConference{XXIIIrd International Symposium on Lattice Field Theory\\
                 25-30 July 2005\\
                 Trinity College, Dublin, Ireland}

\begin{document}
\section{Introduction}
\begin{figure}[t]
  \epsfig{file=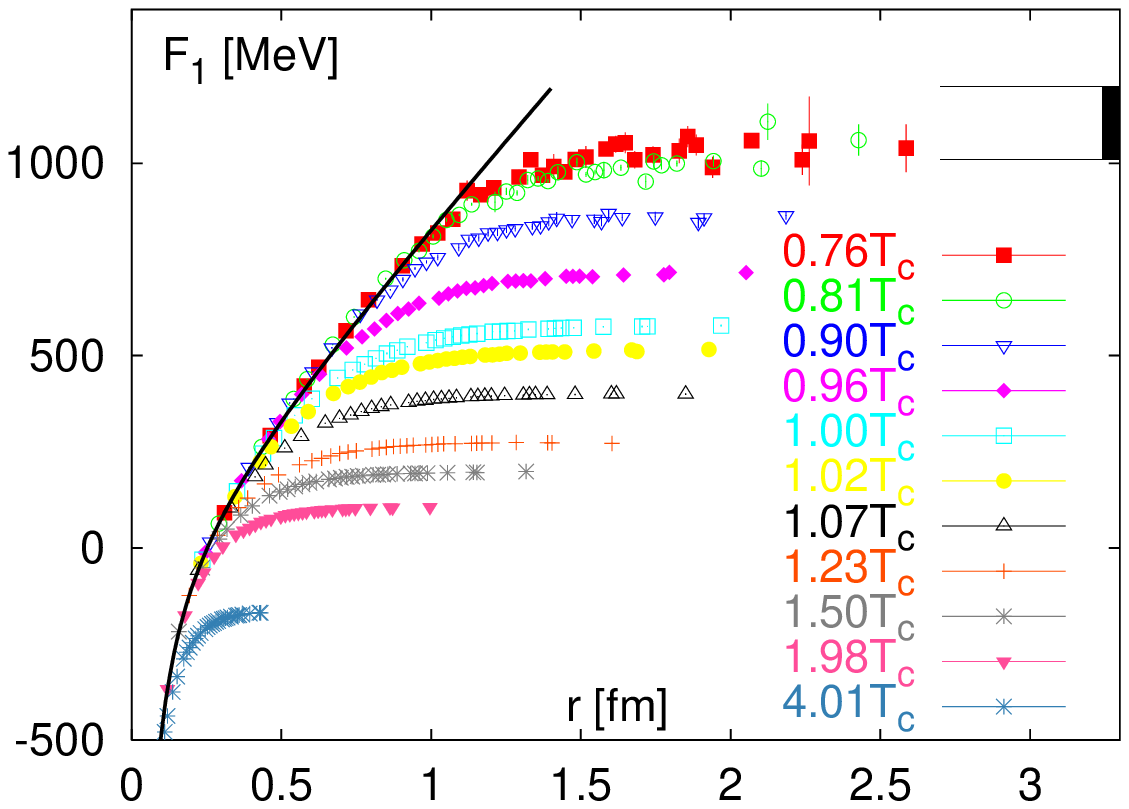,width=7.5cm}
  \epsfig{file=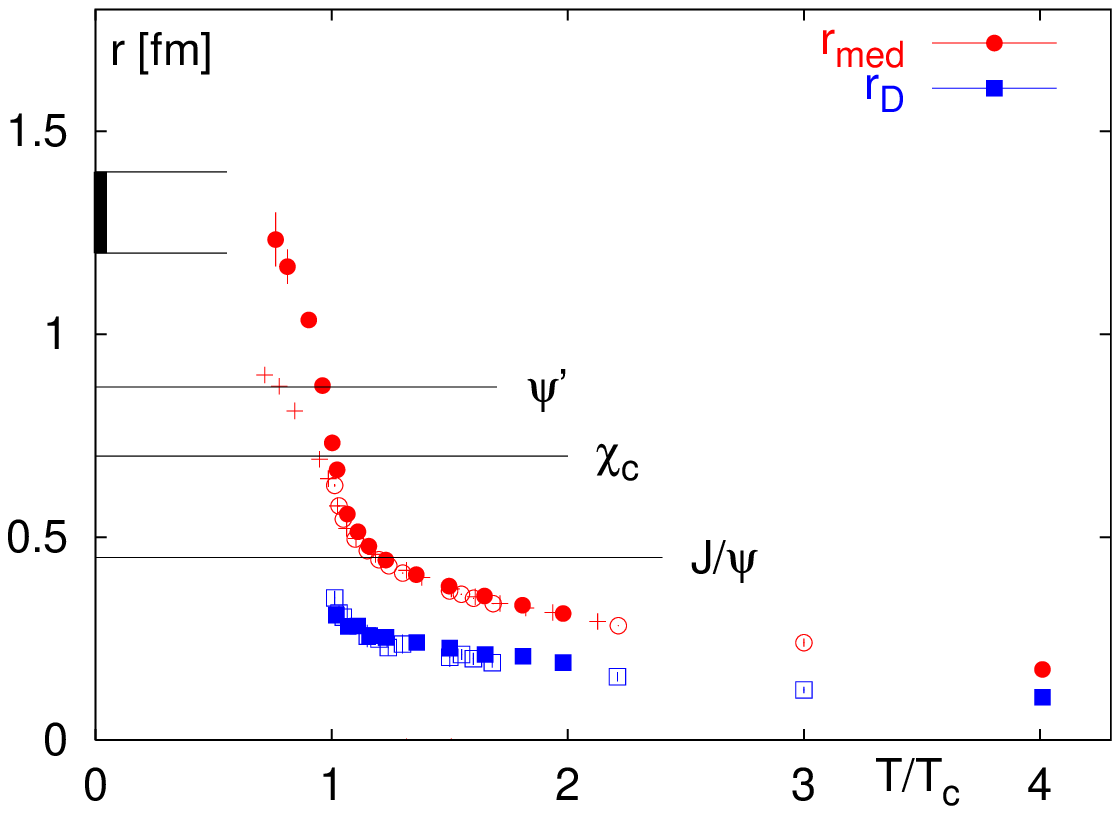,width=7.5cm}
\caption{
  (left) The colour singlet quark anti-quark free energies, $F_1(r,T)$, at several
  temperatures as function of distance in
  physical units. Shown are results from lattice studies of $2$-flavour QCD (from
  \cite{Kaczmarek:2005ui}). The
  solid line represents the $T=0$ heavy quark potential, $V(r)$.
  The dashed error band corresponds to the string breaking energy at zero
  temperature, $V(r_{\text{breaking}})\simeq1000-1200$ MeV, based on the
  estimate of the string breaking distance, $r_{\text{breaking}}\simeq1.2-1.4$
  fm \cite{Pennanen:2000yk}.
  (right)
  The screening radius estimated from the inverse Debye mass, $r_D\equiv1/m_D$
  ($N_f$=0: open squares, $N_f$=2 filled squares), and the scale $r_{med}$
  ($N_f$=0: open circles, $N_f$=2: filled circles, $N_f$=3: crosses) defined in
  (2.1) as function of $T/T_c$. 
  The horizontal lines give the mean squared charge radii of some charmonium
  states, $J/\psi$, $\chi_c$ and $\psi\prime$ (see also
  \cite{Kaczmarek:2005gi,Karsch:2005ex})
  and the band at the left frame shows the distance at which string breaking is
  expected in $2$-flavor QCD at $T=0$ and quark mass $m_\pi/m_\rho\simeq0.7$
  \cite{Pennanen:2000yk}.
}
\label{fesfig}
\end{figure}
A simple Ansatz to study the possible existence of bound states above the critical
temperature is to use effective temperature dependent potentials that model the
medium modifications of strong interactions in a quark gluon plasma. 
To what extend a suitable effective potential at finite temperature can be
defined by quark antiquark free or internal energies and 
furthermore how realistic such (simple) descriptions of bound states in a
deconfined medium are is still an open question.\\
By comparing the screening radii obtained from lattice results on singlet free
energies in 2-flavour QCD to the mean squared charge radii we obtain first estimates on
the temperatures where charmonium bound states may be influenced by medium
effects. In more realistic potential model calculations effective temperature
dependent potentials that model medium effects are used in the Schr\"odinger
equation. We present the heavy quark free energies and their
contributions, i.e. entropy and internal energy, and discuss the different
results obtained using those contributions in potential models.
\section{Screening radii and medium modifications}
In Fig.~\ref{fesfig}~(left) we show results for the heavy quark anti-quark
free energies in 2-flavour QCD \cite{Kaczmarek:2005ui}. 
While in the limit of short distances $F_1(r,T)$ shows no or only little medium
effects, i.e. $F_1(r\rightarrow 0)\simeq V(r)$, at large distances the
free energies approach temperature dependent constant values,
$F_\infty(T)\equiv F_1(r\rightarrow \infty,T)$.
To characterise distances at which medium effects become important we introduce 
a screening radius, $r_{med}$, defined by the distance at which the value of the
zero temperature potential reaches the asymptotic value of the free energies,
\begin{eqnarray}
V(r_{med}) = F_\infty(T).
\label{eqrmed}
\end{eqnarray}
The results for $r_{med}$ are compared to the mean squared charge radii (at zero
temperature) of typical charmonium states in Fig.~\ref{fesfig}~(right). 
The
temperature at which these radii are equivalent can give a rough estimate for
the onset of thermal effects on the charmonium states. Based on this (simple)
picture, $J/\psi$ may survive as a bound state up to temperatures slightly
above the deconfinement temperatures, while $\chi_c$ and $\psi'$ are expected to
show significant thermal modifications already close to the transition.
\section{Asymptotic entropy and internal energy}
\begin{figure}[t]
  \epsfig{file=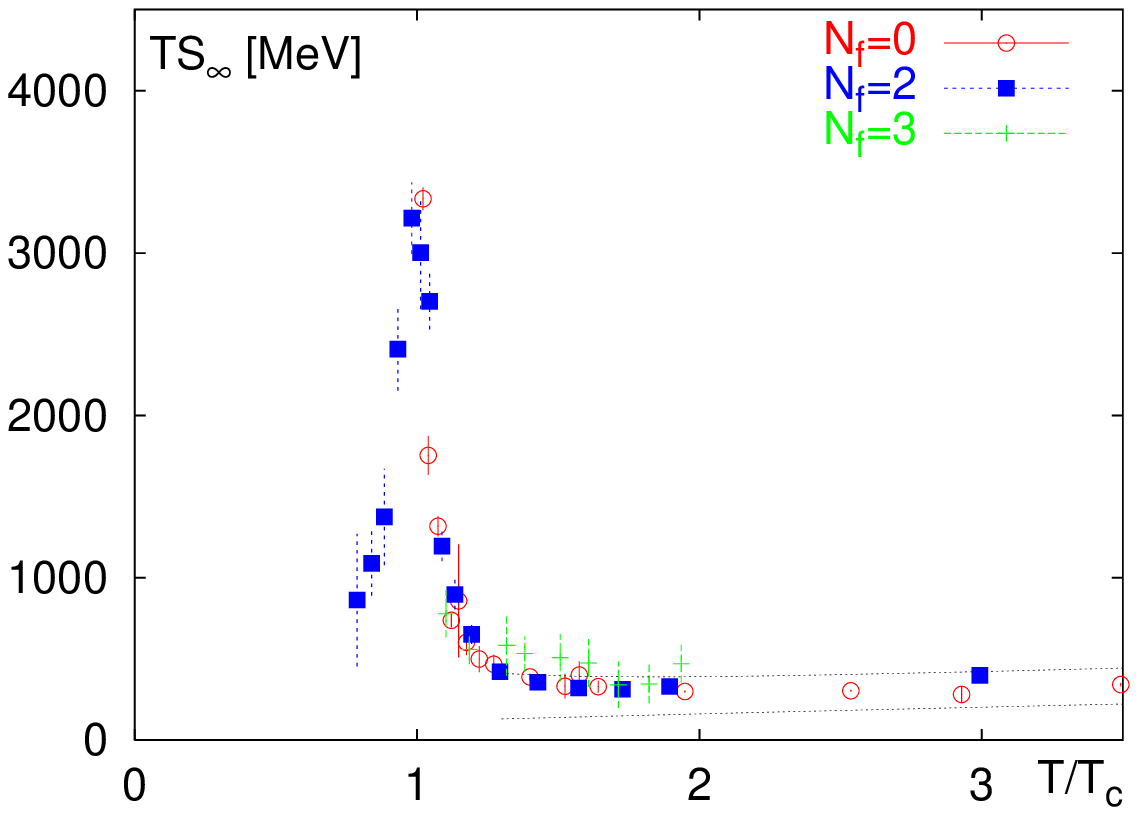,width=7.5cm}
  \epsfig{file=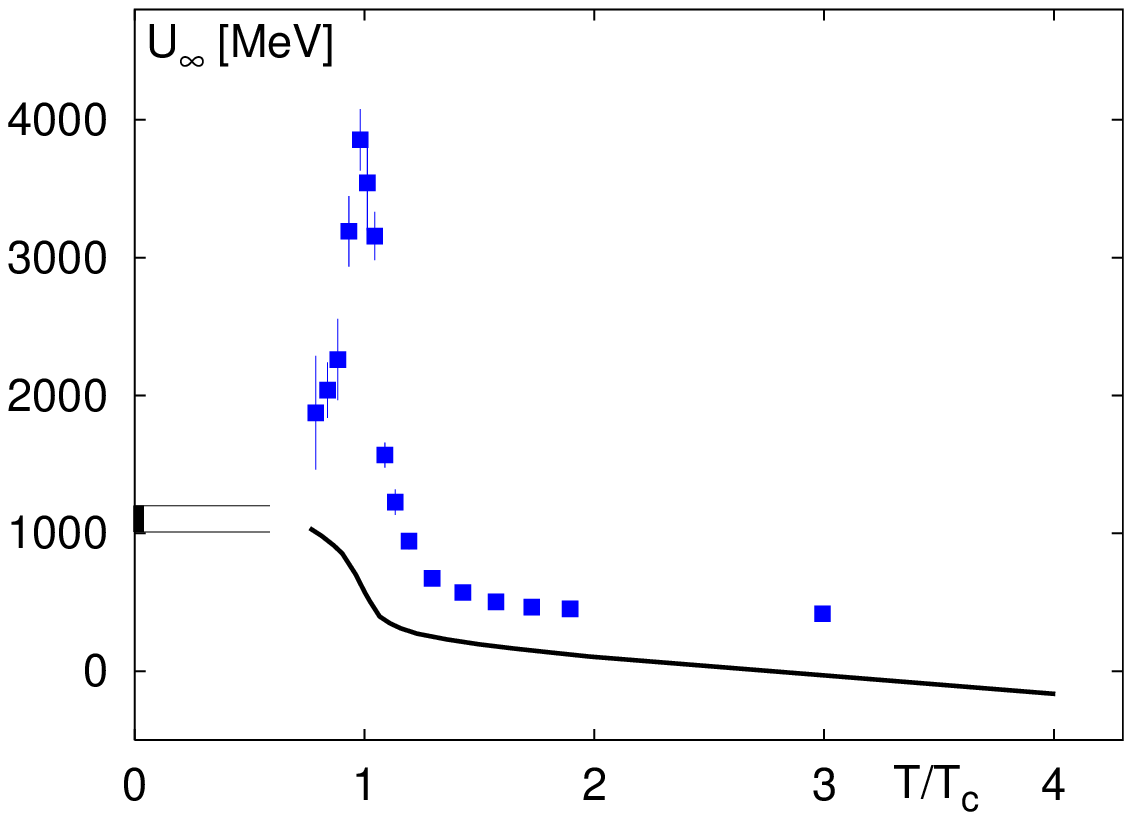,width=7.5cm}
\caption{
  (left)
  The contribution $TS_\infty(T)$ appearing in the free energy,
  $F_\infty(T)=U_\infty(T)-TS_\infty(T)$, calculated in $2$-flavour QCD as
  function of $T/T_c$. We compare our results from $2$-flavour QCD to the
  leading order perturbative result (see \cite{Kaczmarek:2005gi}). We also show results from
  $3$-flavour QCD for $T\;\gsim\;1.1T_c$ \cite{Petreczky:2004pz} and quenched
  QCD \cite{Kaczmarek:2002mc}.
  (right)
  The internal energy $U_\infty(T)$ versus $T/T_c$ calculated in $2$-flavour
  QCD. The corresponding free energy, $F_\infty(T)$, calculated in $2$-flavour
  QCD is also shown as solid line. We again indicate in this figure the energy
  at which string breaking is expected to take place at $T=0$,
  $V(r_{\text{breaking}})\simeq 1000\;-\;1200$ MeV (dashed lines), using
  $r_{\text{breaking}}=1.2\;-\;1.4$ fm \cite{Pennanen:2000yk}. 
}
\label{svfig}
\end{figure}
Before analysing the $r$-dependence of the different contributions to the free
energy, we discuss their behaviour at infinite separation.
The entropy contribution, $T S_\infty(T)$, and internal energy, $U_\infty(T)$
can be calculated from the
asymptotic value of the free energies, $F_\infty(T)$ (see Fig.~2 in
\cite{Kaczmarek:2005gi}) using the thermodynamic relations
\begin{eqnarray}
U_\infty(T) = -T^2\frac{\partial F_\infty(T)/T}{\partial T} \ \ , \ \ \
\ 
S_\infty(T) = -\frac{\partial F_\infty}{\partial T}.\label{us2}
\end{eqnarray}
The results for two flavour QCD compared to quenched and 3-flavour results are
shown in Fig.~\ref{svfig}. The results for small temperatures indicate that
$T S_\infty(T)$ vanishes in the zero temperature limit while at high
temperatures we find a tendency for an increase of $TS_\infty(T)$ with
increasing temperature.
This behaviour is also as expected from leading order perturbative contribution,
\begin{eqnarray}
S_\infty(T) \simeq \frac{4}{3}\frac{m_D(T)}{T}\alpha(T) &+& 4
\frac{m_D(T)}{T}\alpha(T) \frac{\beta(g)}{g(T)},\\
\mathrm{i.e. \ \ \ } T S_\infty(T) &\simeq& {\cal{O}}(g^3 T).
\end{eqnarray}
In contrast to the expected behaviour of the entropy contributions in the limit
of small and
large temperatures, in the vicinity of the phase transition significant
differences are evident. Even at moderate temperatures above $T_c$ the entropy
is to a large extend dominated by non-perturbative effects reaching large
values around the critical temperature. A similar
behaviour is also visible for the internal energies (Fig.~\ref{svfig}~(right))
around $T_c$. A comparison to the free energies (solid line) shows that
$U_\infty(T) > F_\infty(T)$ at all temperatures analysed here. We stress here
that even at 
high temperatures the difference between both show that entropy
contributions play an important role. At small
temperatures, the comparison to the expected value at zero temperature
indicates that this value seems to be approached already at rather large
temperatures. This is in agreement with the observation that only small
temperature effects are visible in the free energies in
Fig.~\ref{fesfig}~(left) below $0.8~T_c$.
\section{$r$-dependent entropies and internal energies}
\begin{figure}[t]
  \epsfig{file=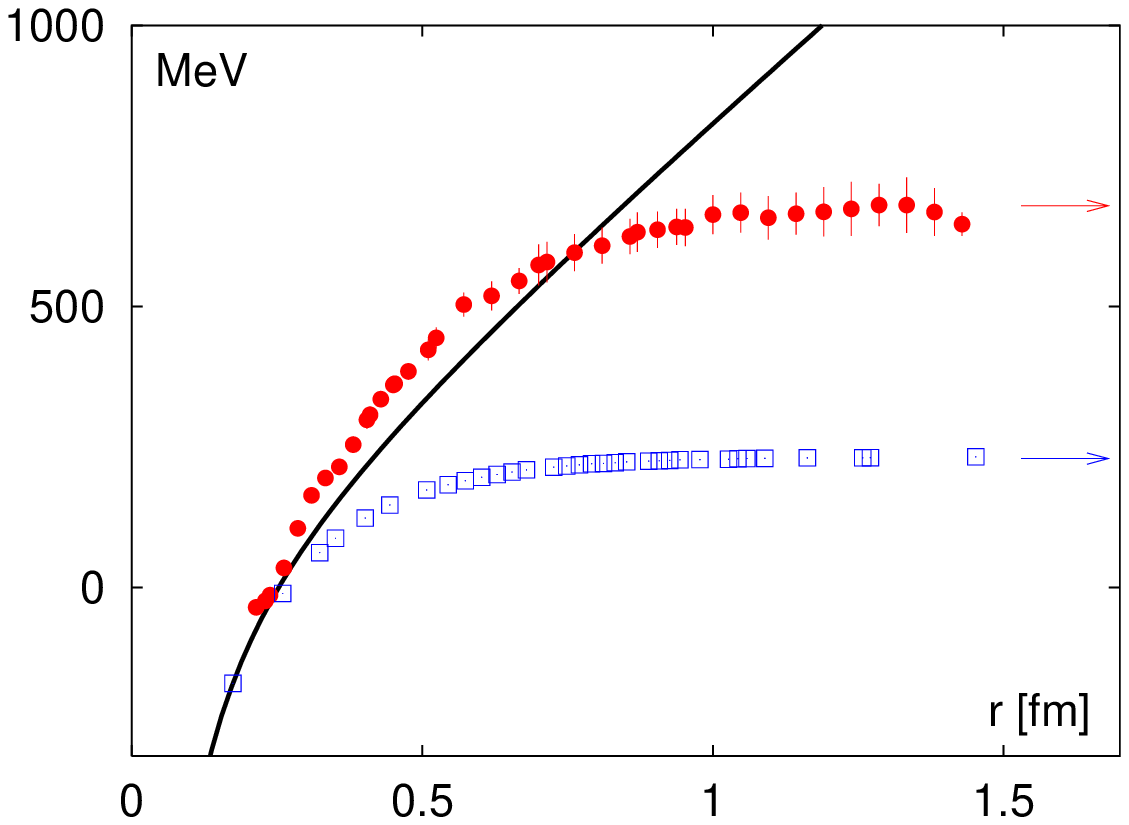,width=7.5cm}
  \epsfig{file=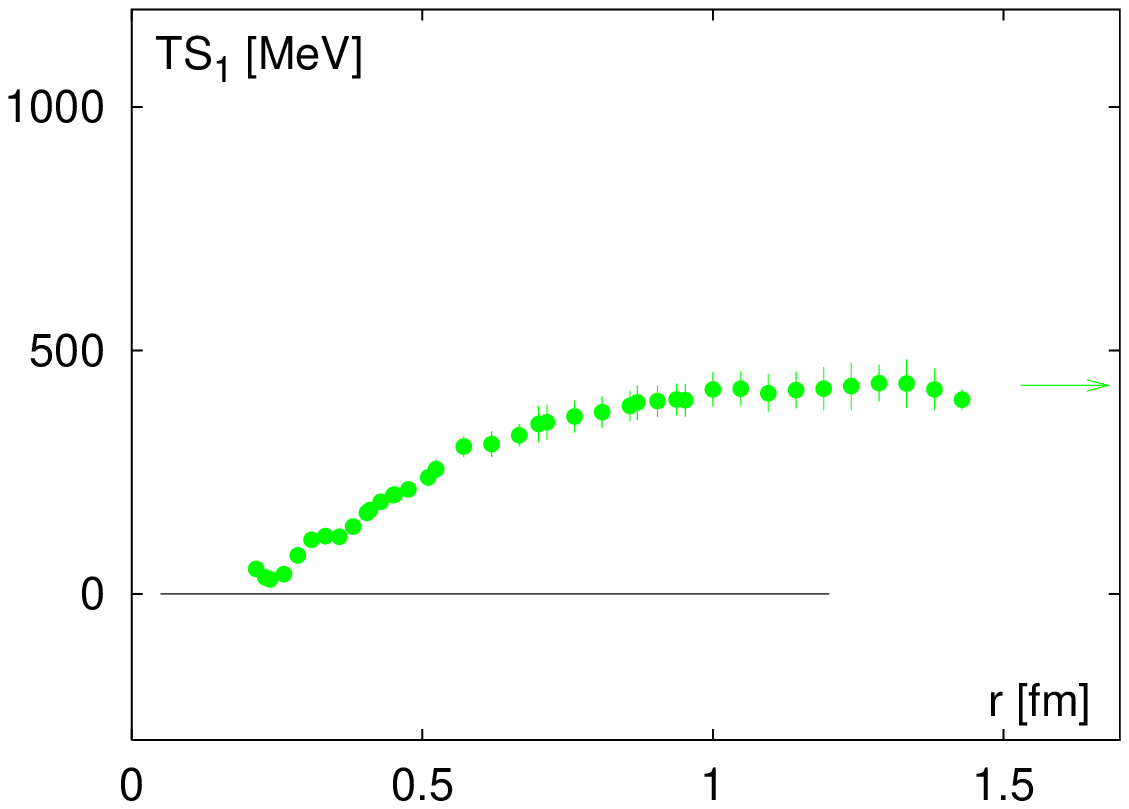,width=7.5cm}
\caption{
  (left) The singlet internal energy, $U_1(r,T)$ (filled circles),
  calculated from renormalised singlet free energy, $F_1(r,T)$ (open squares),
  at fixed $T\simeq1.3T_c$ in $2$-flavour lattice QCD compared
  to $V(r)$ (line) \cite{Kaczmarek:2005ui,Kaczmarek:2005uw}.
  (right) 
  The
  corresponding colour singlet quark anti-quark entropy,
  $TS_1(r,T\simeq1.3T_c)$, as function of distance calculated from renormalised
  free energies. The arrows in both figures point at the temperature dependent
  values of the free and internal energy and entropy at asymptotic large
  distances, {\em i.e.} $F_\infty(T)\equiv \lim_{r\to\infty}F_1(r,T)$,
  $U_\infty(T)\equiv \lim_{r\to\infty}U_1(r,T)$ and $TS_\infty(T)\equiv
  T\lim_{r\to\infty}S_1(r,T)$.
}
\label{pot1fig}
\end{figure}
\begin{figure}[t]
  \epsfig{file=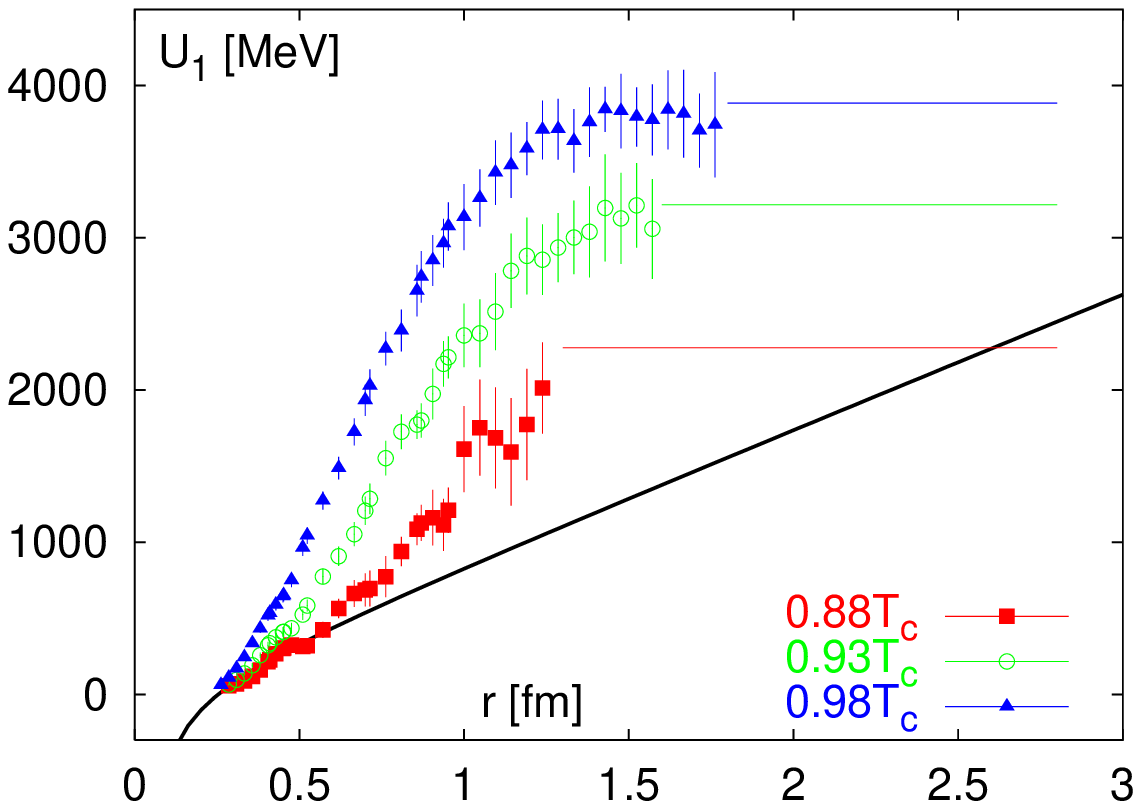,width=7.5cm}
  \epsfig{file=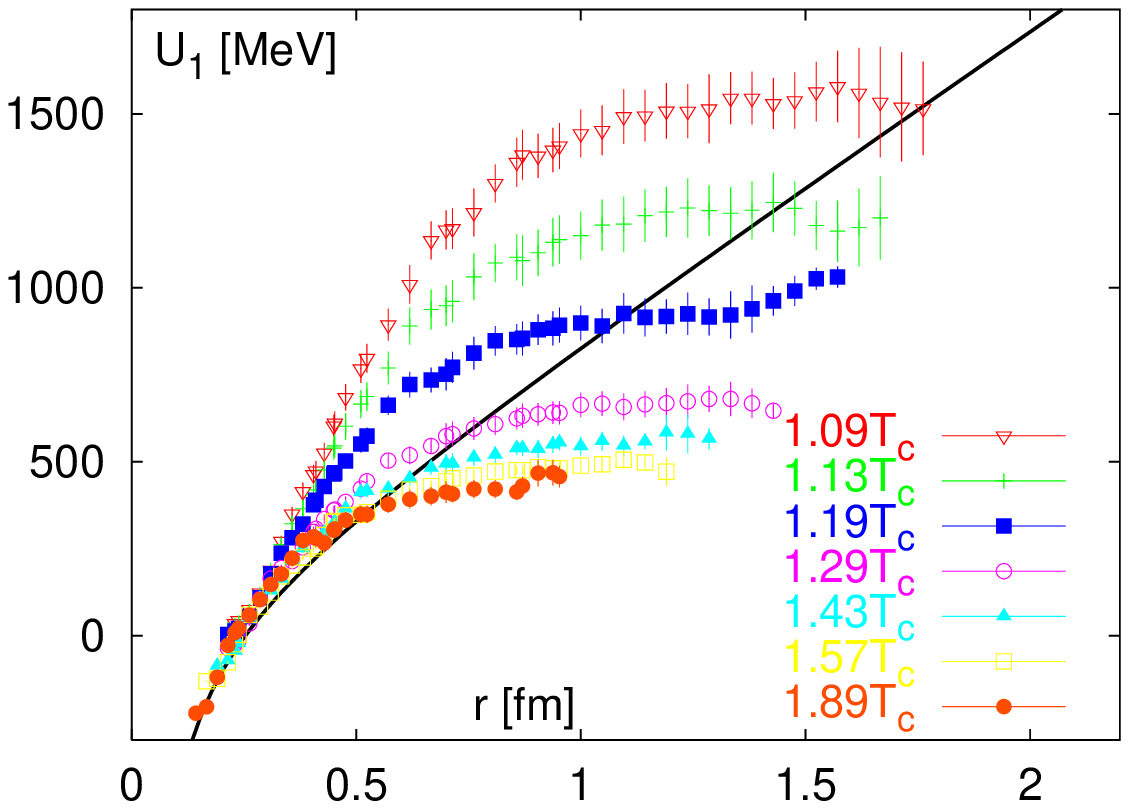,width=7.5cm}
\caption{
  The colour singlet quark anti-quark internal energies, $U_1(r,T)$, at
   several temperatures below (left) and above (right) the phase transition obtained
   in $2$-flavour lattice QCD. In (left) we also show as horizontal lines the
   asymptotic values which are approached at large
   distances and indicate the flattening of $U_1(r,T)$.
   The solid lines
   represent the $T=0$ heavy quark potential, $V(r)$
   \cite{Kaczmarek:2005ui,Kaczmarek:2005uw}.
}
\label{pots_paper}
\end{figure}

We now turn to the discussion of the $r$-dependent entropy and internal energy
contributions calculated by the appropriate relations as (\ref{us2}) and
(\ref{us2}). In Fig.~\ref{pot1fig}~(left) the internal energy at a temperature
$T\simeq 1.3~T_c$ is compared to the free energy. Both energies approach the
zero temperature potential at small distances. Therefore the free energy at
small separations is dominated by energy contributions. The results of
$T S_1(r,T)$ in Fig.~\ref{pot1fig}~(right) indicate that at intermediate and
large distances entropy contributions play an important role leading to an
enhancement of the internal energy compared to the free energy. In
Fig.~\ref{pots_paper} we summarise results for $U_1(r,T)$ at various temperatures
below (left) and above (right) $T_c$. The results for $U_1(r,T)$ are larger
than $F_1(r,T)$ for all temperatures
and show a much steeper slope compared to $F_1(r,T)$. Therefore potential
models using $U_1(r,T)$ as an effective $T$-dependent potential will lead to
stronger bound states compared to models using free energies as effective
potential.
\section{Bound states in potential models}
Various potential model calculations in terms of solving the Schr\"odinger
equation using either free energies \cite{Digal:2001ue}, internal
energies \cite{Alberico:2005xw} or a linear combination of both
\cite{Wong:2005be} 
were recently performed leading to
different results for the temperature dependence of binding energies of heavy
quark bound states in the quark gluon plasma. Some quarkonium dissociation
temperatures obtained by assuming vanishing binding energy are summarised in
Tab.~\ref{tab1}. Although the results differ, with the smallest dissociation
temperatures obtained using $F_1(r,T)$ and the highest using $U_1(r,T)$, they
indicate that at least $J/\psi$ may survive the deconfinement transition as a
bound state, while the situation for the higher states is still not obvious.
\begin{table}[thbp]
\centering
\begin{tabular}{|c||c|c|c||c|c|c|c|c|}
\hline
state & $J/\psi$ & $\chi_c$ & $\psi'$ & $\Upsilon$ & $\chi_b$ & $\Upsilon'$ &
$\chi'_b$ & $\Upsilon''$\\
\hline
$E_s^i [GeV]$ & 0.64 & 0.20 & 0.005 & 1.10 & 0.67 & 0.54 & 0.31 & 0.20\\
\hline
{\color{green} $T_d/T_c$} & {\color{green}1.1} & {\color{green}0.74} &
{\color{green}0.1-0.2} & {\color{green}2.31} & {\color{green}1.13} &
{\color{green}1.1} & {\color{green}0.83} &
{\color{green}0.75}\\
{\color{blue} $T_d/T_c$} & {\color{blue}$\sim 1.42$} & {\color{blue}$\sim 1.05$}
& {\color{blue}unbound} & {\color{blue}$\sim 3.3$} &
{\color{blue}$\sim 1.22$} & {\color{blue}$\sim 1.18$} & {\color{blue}-} & {\color{blue}-}\\
{\color{red} $T_d/T_c$} & {\color{red} 1.78-1.92} & {\color{red} 1.14-1.15}
& {\color{red} 1.11-1.12} & {\color{red}$\gsim$4.4} &
{\color{red} 1.60-1.65} & {\color{red} 1.4-1.5} & {\color{red}
  $\sim 1.2$} & {\color{red} $\sim 1.2$}\\
\hline
\end{tabular}
\caption{
Estimated dissociation temperatures $T_d$ in units of $T_c$ obtained from
potential models using free energies \cite{Digal:2001ue} (green), a linear combination
of $F_1$ and $U_1$ \cite{Wong:2005be} (blue) and internal energies
\cite{Alberico:2005xw} (red)
as effective $T$-dependent potentials.
}
\label{tab1}
\end{table}
\section{Conclusions}
We have compared the screening radii of heavy quark anti-quark pairs in the
quark gluon plasma phase to the (zero temperature) mean squared charge radii of
charmonium states and found indications that the $J/\psi$ may survive the phase
transition as a bound state, while $\chi_c$ and $\psi'$ are expected to show
significant thermal modifications at temperatures close to the transition.\\
Beyond this simple approximation of the medium modifications of charmonium
bound states above $T_c$, we have also analysed the different contributions to the
heavy quark free energy and calculated the entropy contributions and internal
energy of a static quark anti-quark pair. A comparison of different potential
models, using either free energies, internal energies or a combination of both,
shows that charmonium states may survive the phase transition and exist as bound
states in the quark gluon plasma. 
This is in (qualitative) agreement with spectral function analyses in quenched QCD 
\cite{Asakawa:2002xj,Asakawa:2003re,Datta:2003ww} and first
results obtained in 2-flavour QCD \cite{Morrin:2005zq}.
As the systematic uncertainties in all those analyses are still quite large,
up to now it is not clear to which
temperatures bound states may exist and which potential models give a realistic
description of charmonium or bottomonium systems at high temperatures.
Clearly more detailed calculations beyond these simple potential models are
needed to clarify the possibility of bound states in the quark gluon plasma as
well as their medium modifications and dissociation properties in a deconfined
medium. A comparison of the various potential models to (direct) lattice
calculations of charmonium correlation and spectral functions may clarify the
question if and which potential models lead to the correct description of bound
state phenomena within their applicability.
\subsection*{Acknowledgements}
We would like to thank E.~Laermann and F.~Karsch for many fruitful
discussions. F.Z. thanks P.~Petreczky
for his continuous support. This work has partly been supported by DFG under
grant FOR 339/2-1 and by BMBF under grant No.06BI102 and partly by contract
DE-AC02-98CH10886 with the U.S. Department of Energy. At an early stage of this
work F.Z. has been supported through a stipend of the DFG funded graduate
school GRK881.

\end{document}